\documentclass[runningheads]{llncs}
\usepackage{graphicx}
\usepackage{multicol}
\usepackage{multirow}
\usepackage{xcolor}
\usepackage{comment}

\begin{document}
\title{Towards Stroke Patients' Upper-limb Automatic Motor Assessment Using Smartwatches} 
\titlerunning{Stroke Patients' Upper-limb Automatic Motor Assesment}

%
%
\author{Asma Bensalah\inst{1}\orcidID{0000-0002-2405-9811} \and
Jialuo Chen\inst{1}\orcidID{0000-0002-7808-6567} \and
Alicia Forn{\'e}s\inst{1}\orcidID{0000-0002-9692-5336} \and
Cristina Carmona-Duarte\inst{2}\orcidID{0000-0002-4441-6652} \and Josep Llad{\'o}s\inst{1}\orcidID{0000-0002-4533-4739} \and Miguel A.Ferrer\inst{2}\orcidID{0000-0003-4913-4010}
}
\authorrunning{A. Bensalah et al.}
%
\institute{Computer Vision Center, Computer Science Department, \\Universitat Aut\`{o}noma de Barcelona, Spain\\
\email{\{abensalah, jchen, afornes, josep\}@cvc.uab.es}\\
 \and
Instituto Universitario para el Desarrollo Tecnol{\'o}gico y la Innovaci{\'o}n en Comunicaciones, Universidad de Las Palmas de Gran Canaria, Spain\\
\email{\{ccarmona\}@idetic.eu}}
\maketitle              

\begin{abstract}
Assessing the physical condition in rehabilitation scenarios is a challenging problem, since it involves Human Activity Recognition (HAR) and kinematic analysis methods. In addition, the difficulties increase in unconstrained rehabilitation scenarios, which are much closer to the real use cases. In particular, our aim is to design an upper-limb assessment pipeline for stroke patients using smartwatches. We focus on the HAR task, as it is the first part of the assessing pipeline. Our main target is to automatically detect and recognize four key movements inspired by the Fugl-Meyer assessment scale, which are performed in both constrained and unconstrained scenarios. In addition to the application protocol and dataset, we propose two detection and classification baseline methods. We believe that the proposed framework, dataset and baseline results will serve to foster this research field.
\end{abstract}


%

\section{Introduction} \label{s:introduction}

Neuromuscular diseases (e.g. multiple sclerosis, Parkinson's disease) and strokes (cerebrovascular accident) involve a loss in the motor control system. In the particular case of stroke patients, the rehabilitation stage is crucial for minimizing, as much as possible, their deficits or motor disabilities towards their social reintegration. During the rehabilitation process, which mainly consists in drug therapy and rehabilitation exercises, the patient's neuromotor condition and progress must be evaluated. But above all, a continuous and accurate estimation is necessary during the early rehabilitation stages (e.g. first weeks) so that the neurologist can monitor the patient's improvement and adapt the rehabilitation therapy (e.g. modify the medication doses) before the patient's motor impairment becomes irreversible (the most notable progress is usually achieved during the first weeks). Unfortunately, the probability that the patient fully recovers the upper-limb mobility is very low ($<15\%$)~\cite{medical}. 


Monitoring the patient's neuromotor conditions involves Human Activity Recognition (HAR) tasks, preferably in a continuous unconstrained scenario. Methods for HAR use input data from video images \cite{zunino2018video} or from time series signals acquired  with  on-body  sensors \cite{rueda2018learning}. In the last fourteen years, most HAR systems have either focused on entire action recognition within a constrained scenario~\cite{non-repetitive}~\cite{limb} or repetitive movements in an unconstrained scenario~\cite{recofit}. HAR in a constrained scenario refers to recognizing an action among a set of, solely, well defined actions (namely, target actions), in terms of action's content and performance style. Contrary, HAR in an unconstrained scenario means recognizing an action in a melange of well defined actions (like in a constrained scenarios) that are performed together with other actions either loosely defined or not defined at all (namely, non-target actions).

An action in HAR is a composition of body movements. In that sense, actions split into two categories: Actions with repetitive movements (e.g. walking/running are actions that involve the lifting foot movement repetitively), and actions with non-repetitive movements (e.g. grabbing something). Although it is difficult to segment actions and therefore recognize them in an unconstrained scenario, it is even more difficult when the (target) actions to recognize are compound of non-repetitive movements. The main reason is that the repetition can serve as context (like objects vs background in an image), hence, the classification and segmentation are context-aware. However, that context is absent in non-repetitive movements scenarios. This is the case of the upper-limb movements used for the Fugl-Meyer assessment (FMA)~\cite{Juan-carlos-univ}, one of the most frequently used metric scales for stroke patients.



Therefore, our motivation is to design an upper-limb assessment framework for stroke patients. This work has been developed in the context of the \emph{3D kinematics for remote patient monitoring} (RPM3D) project\footnote{\url{http://dag.cvc.uab.es/patientmonitoring/}}. The main goal of this project is to derive an objective estimator of the improvement of the patients’ motor abilities during rehabilitation through the analysis of the 3D movements captured with smartwatches (worldwide affordable and non-intrusive technology). On the basis of the context of our study and the target population, we opted for "non-intrusive" sensors. Our choice went to Apple Watch Series 4, which is an FDA-cleared class 2 medical device and less expensive than existing high-end clinical devices. Moreover, in order to deploy a real assessment on the stroke patient's stated amid the rehabilitation process, we have designed an unconstrained data experimentation scenario, similar to the real conditions outside of a lab or therapy rooms, towards a continuous (24/7) patient monitoring. 

Within this general objective in mind, in this paper we focus on the first part of this pipeline. We have simulated in the lab the gesture capture with smartwatch and the subsequent analysis conditions for stroke patients. This framework is exportable to 24/7 patient monitoring in daily life conditions. Thus, the first contribution of this work is the design of an evaluation protocol based on the Fugl-Meyer scale for constrained and unconstrained scenarios with non-repetitive movements. It consists of a set of target actions acquired from the movements of a study population. We got inspired from the FMA to outline the well defined target actions, to which we appended a set of loose (non-target) defined actions. In this way we simulate a real use case scenario, in which the patient wears the smartwatch all day (so, continuous recording). The dataset contains samples from healthy subjects and stroke patients and has been manually annotated. The dataset will be made available for public use to foster the HAR research for stroke rehabilitation purposes. 


The second contribution of this work consists of an Activity Recognition Chain (ARC) for detecting and classifying gestures in the constrained and unconstrained FMA inspired scenario. Since the input data recorded by the sensors is a continuous time series data, actions must be detected as subsequences. We propose two segmentation approaches: in the first one, namely \emph{action segmentation}, the subsequence covers the entire action, whereas in the second one, namely \emph{gesture spotting}, the subsequence only covers a part of it (the gesture). We also propose two classification methods: the first one is based on Support Vector Machines (SVM), whereas the second is based on  Convolutional Neural Networks (CNN). These methods can serve as baseline results. The correct detection and segmentation of these subsequences is important, so that they can be properly analyzed by the kinematic model~\cite{Schindler2018ExtendingTS}, estimating the improvement of the neuromotor control system of the patient. 


The rest of this paper is organized as follows. In section \ref{s:SoA} we overview the existing methods related to our work. Section \ref{s:ESetup} describes the experimentation protocol, based on the Fugl-Meyer Assessment. Section \ref{s:method} describes our methodology in detail, including the data capturing, preprocessing and the gesture spotting. Section \ref{s:Results} shows the experimental baseline results, and Section \ref{s:Conclusions} draws the conclusions and future work.

\section{State of the Art} \label{s:SoA}


Efficient methods for HAR have to tackle with the traditional pattern recognition problems, namely intraclass variability (performance differences within the same individual) and interclass similarities (performance similarities between individuals)~\cite{interclass}. When gesture spotting is faced, the NULL class problem arises. When the objective is to spot a certain number of prototype (target) activities in the input signal, the rest of non-spotted (non-target) activities fall in the NULL class. The diversity in the NULL class makes it difficult to model. Moreover, there are specific issues related to the nature of the activity recognition problem and the data itself, for instance, the loose definition of a physical activity, the data labelling or the experiment design and setup \cite{Tutorial}.


Clarkson Patrick \cite{clarkson2002life} approached the feasibility of computationally structuring human daily activities, based on the raw sponsors' data. In addition, they addressed the challenging task of structures' similarity, perplexity, prediction and classification. The picked sensors aim to reproduce the natural insect senses: the eyes (two cameras), the hearing (microphone) and gyros for orientation. 
As an extension of this work, a first algorithm that explores the efficiency of human activity recognition algorithm deployed with five biaxial accelerometers was proposed in \cite{Ling}. Following the above cited works, HAR witnessed an enhance in the number contributions, mainly focusing on the segmentation and recognition tasks.

The signal segmentation is applied before the recognition step. This segmentation step is often addressed while designing the experimentation scenario. For example, for facilitating the ground-truthing and labelling process, users were asked to stand still for five seconds in~\cite{segmentation}. Other works use Discrete wavelet transforms to segment the data~\cite{wavelet} and split the input signal into approximations. One of the most perpetual approaches to face annotation scarcity in HAR is sliding  windows~\cite{sliding_window}\cite{U-Net}\cite{bag-words}. A small sliding window can discard crucial information, while larger ones can contain action transactions. Thus, ascertaining the sliding window size is a vital step for bettering off an ARC. 
In another work~\cite{cnn}, Convolutional Neural Networks (CNN) are employed to attribute each label timestep, instead of labeling an entire sliding window~\cite{sliding_window}. 

Concerning the recognition task, several methods were proposed. The Support Vector Machine's (SVM) efficiency and ease incited its wide use in HAR applications~\cite{Ahmed2020EnhancedHA}\cite{SVM}\cite{SVM1}. SVM performs the classification task via separating classes by linear decision border (hyperplane) in the feature space~\cite{svm_linear}.
In tandem, sparse signal representations have also been popular in HAR~\cite{sparse}\cite{sparse1}. Other statistical learning algorithms were broadly used, like Random forest~\cite{random_forest}\cite{random_forest1}.
In the interest of extracting more robust and scalable features to improve the recognition task, deep learning was introduced to HAR~\cite{deep}~\cite{deep1}.

Clearly, one of the HAR field applications is health \cite{Tian}\cite{recofit}, inter alia Neurorehabilitation. Cognate to our work study case, in~\cite{limb} an ARC was implemented to track long-term tremor activities and the treatment effect. In~\cite{stroke} authors provided a proof-of-principle regarding the identification of a set of daily life activities within stroke patients. In~\cite{eloy}, inertial sensors from high-end clinical devices were used for evaluating the functional improvement of stroke patients (in lieu of relying on diaries and self-questionnaire which does not bespeak the patient's real condition). Anyway, the above methods present several limitations: in\cite{recofit} and \cite{Tian}, the data was collected from perfectly healthy subjects, while in \cite{eloy} the work's aim was to classify the movements into purposeful and non-purposeful solely movements rather than recognizing them. 

In summary, and from the state of the art review, we observe that there is still the need to design suitable HAR approaches for Neurorehabilitation in unconstrained and continuous scenarios using affordable and consumer electronic devices such as smartwatches.




\section{Experimentation Protocol} \label {s:ESetup}

This section is devoted to describe our experimental setup. As explained earlier, we are focusing in non-repetitive actions inspired by the Fugl-Meyer Assessment scale, an index to assess the sensorimotor impairment (i.e. the motor functioning, balance, sensation and joint functioning) in stroke patients. Concretely, we have defined four target (or key) movements $\mathcal{M}_i$ based on the following joint movements:
\begin{itemize}
    \item Movement $\mathcal{M}_1$: shoulder extension/flexion.
    \item Movement $\mathcal{M}_2$: shoulder abduction/abduction.
    \item Movement $\mathcal{M}_3$: external/internal shoulder rotation.
    \item Movement $\mathcal{M}_4$: elbow flexion/extension
\end{itemize}

\begin{figure}[t]
  \centering
  \begin{tabular}{cccc}
    \includegraphics[width=0.23\textwidth]{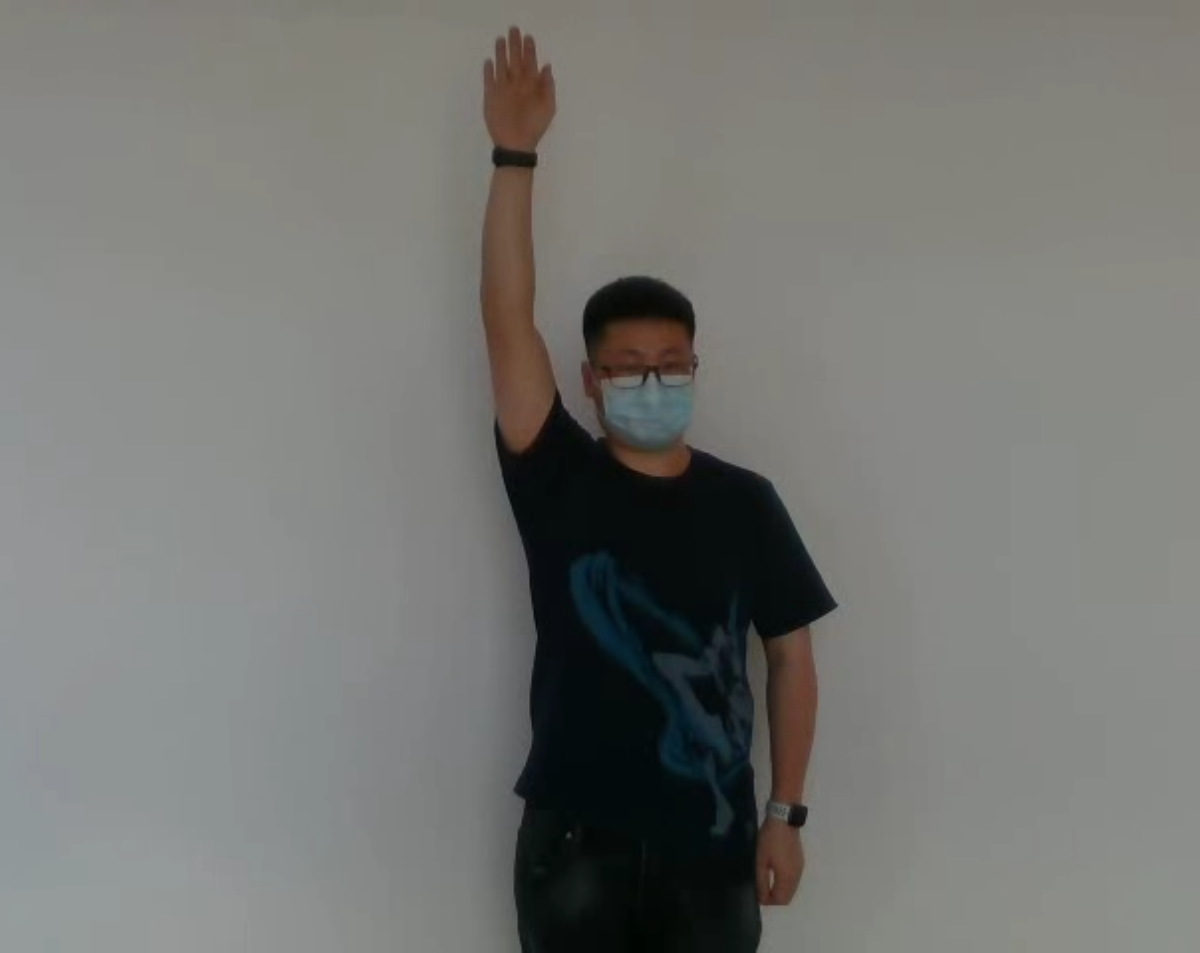} &
    \includegraphics[width=0.23\textwidth]{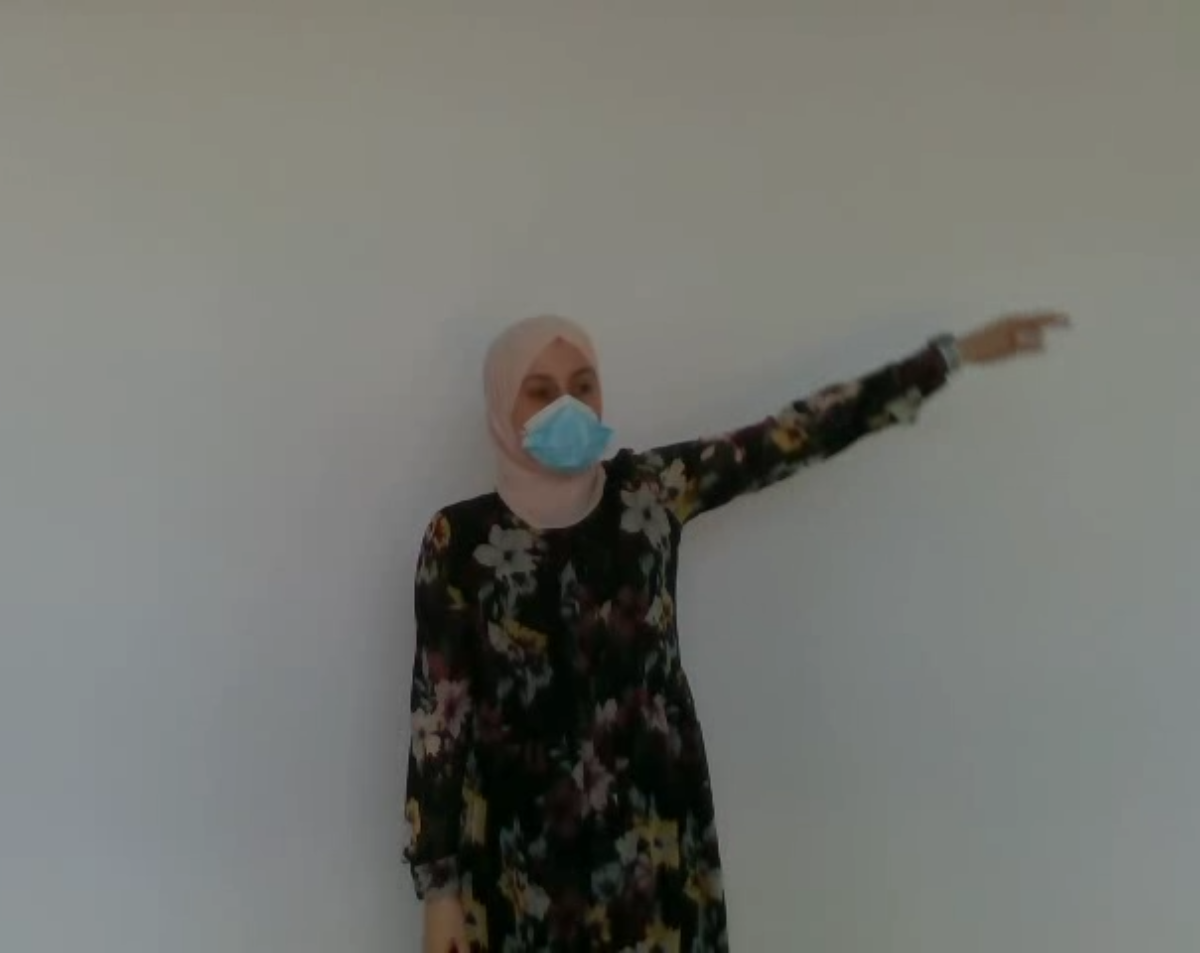} &
    \includegraphics[width=0.23\textwidth]{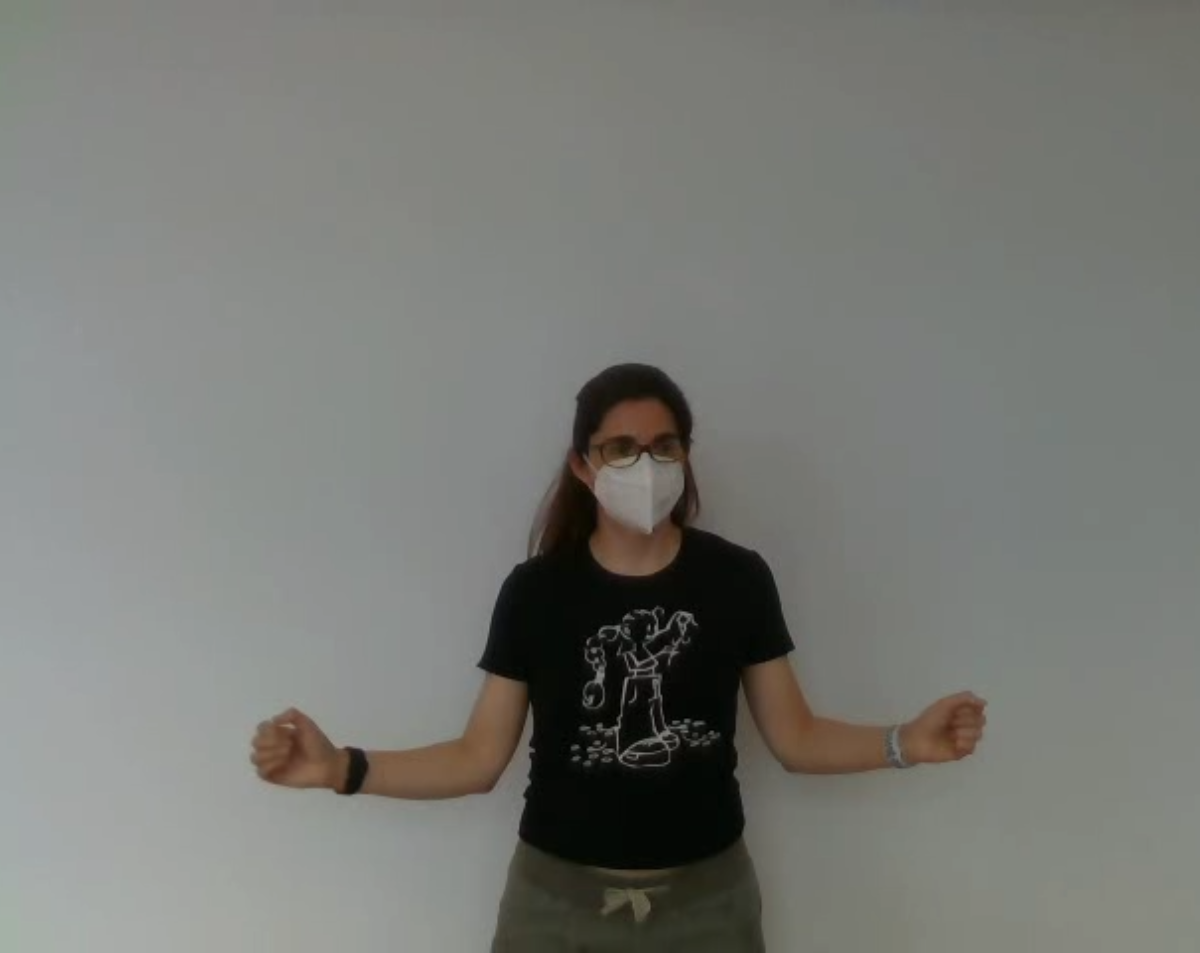} &
    \includegraphics[width=0.23\textwidth]{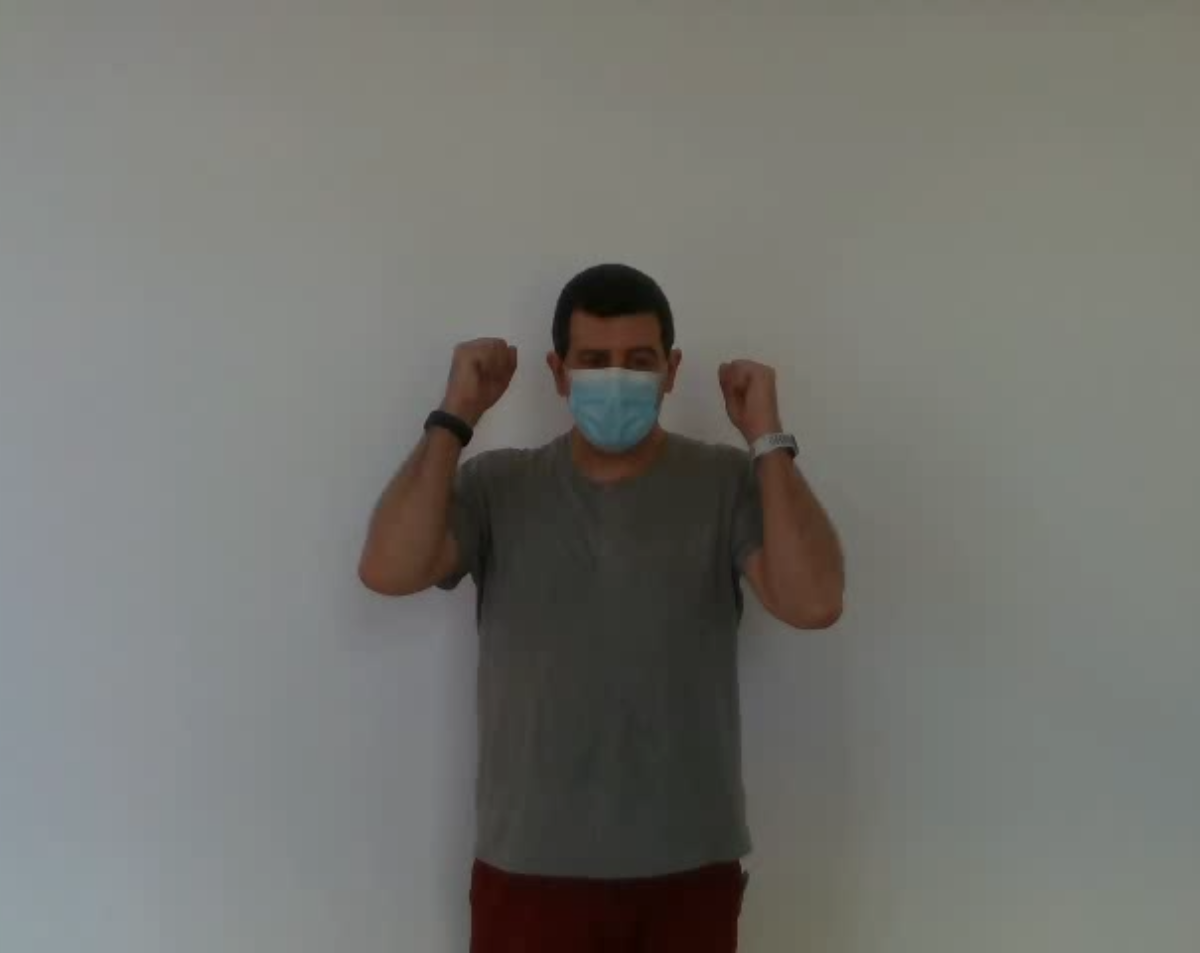} \\
  a) & b) &  c) & d) \\
  \end{tabular}\\
  \caption{Target (key) movements based on the Fugl-Meyer Assessment. a) Movement 1; b) Movement 2; c) Movement 3; d) Movement 4. }
    \label{fig:movements}
\end{figure}

These movements are illustrated in Figure~\ref{fig:movements}.
In order to analyze the performance of the detection and classification methods, we have recorded these movements both in constrained and unconstrained scenarios. Thus, we designed two scenarios, namely L1 and L2, as follows:

\begin{itemize}
\item Scenario L1: It is a constrained scenario which consists in performing the same type of target movement in a sequence, but alternating the arm (left, right or both). Thus, the user performs the movement $\mathcal{M}_i, i\in \left [ 1,4 \right ]$, as follows:
\begin{enumerate}
\item Perform movement $\mathcal{M}_i$ with the dominant hand;
\item Perform movement $\mathcal{M}_i$ with the non-dominant hand;
\item Perform movement $\mathcal{M}_i$ with both hands, simultaneously;
\end{enumerate}

Between each movement, the user is asked to rest calm for 5 seconds. 

\item Scenario L2:
It is an unconstrained scenario, in which the user is performing target $\mathcal{M}_i$ movements in between of longer sequences of non-target $\mathcal{R}_j$ movements. It requires carrying out a movement $\mathcal{M}_i$, $i\in \left [ 1,4 \right ]$, along with with a movement $\mathcal{R}_j$, $j\in \left [ 1,19 \right ]$. The movements $\mathcal{R}_j$ have a loose definition in terms of the action content and the performing style (so they could be seen as background/noise), although they are common daily life activities. Examples of these kind of non-target but realistic movements include: eating, pouring water into a glass, drinking, brushing your teeth, aiming to an object with your arm, getting up, sitting on a chair, applauding, scratching the ear/shoulder, etc. In order to mimic real world conditions, we randomly alternate between target/key $\mathcal{M}_i$ movements and non-target $\mathcal{R}_j$ ones.
\end{itemize}

As it can be noted, the scenario L2 is more difficult than the L1 one  because the sequence of movements is completely random (both the target and non-target ones), so the system can not benefit from information on previous actions.




\section{Methodology} \label{s:method}

The first step of our ARC consists in the data capturing using two smartwatches, one wrist each. This data is recorded in the smartwatch and sent to the mobile phone and the cloud service. The following step consists in the preprocessing and segmentation. The action is classified either as a whole (sliding window), or only analyzing the relevant trimmed parts. The classification task is accomplished via an SVM or a CNN model. 
All these steps are described next.

\subsection{Data capture and preprocessing} 

The data collection consists of recording sequences of movements while wearing two Apple Watch 4 (series 4), one on the left wrist and another one on the right wrist. We record data in both arms because stroke patients usually have one side of the body more affected than the other. The user-generated acceleration (without gravity) for all three axes of the device, unbiased gyroscope (rotation rate), magnetometer, altitude (Euler angles) and temporal information data have been recorded at 100Hz sampling rate and labelled in the smartwatch’s internal memory. Once the data has been recorded, it is transmitted to the mobile phone and the cloud service.

\begin{figure}[h]
\centering
\includegraphics[scale=0.06]{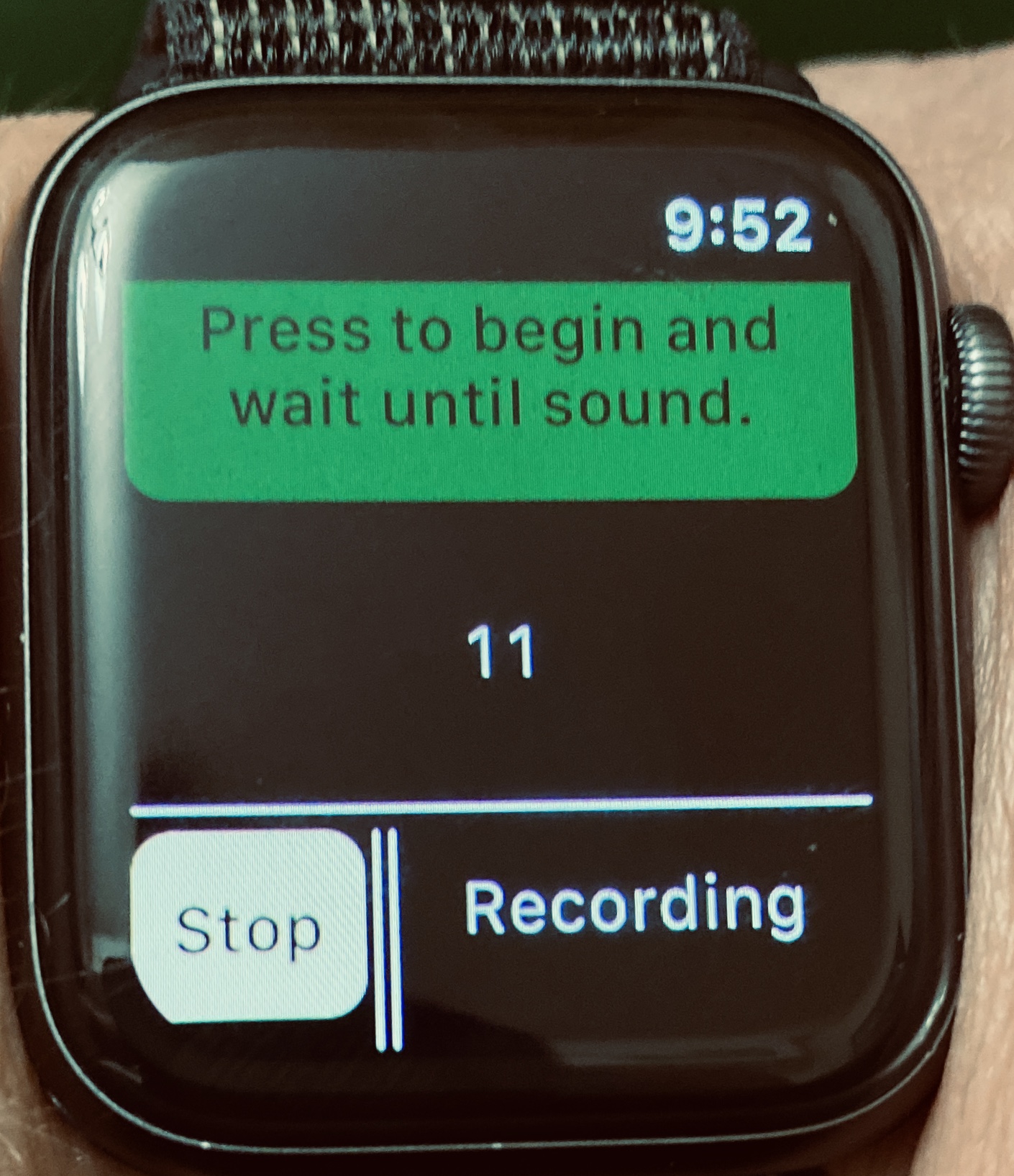} 
\caption{Developed smartwatch application for data capturing.}
\label{fig:app}
\end{figure}

We have developed an application for the smartwatch, as shown in Figure~\ref{fig:app}, that allows selecting the recording time and the user's number. Once the user is ready to start recording, he/she can tap the corresponding button. When the two smartwatches are synchronized, they emit an audio and visual signal to inform that the recording has started. In this way, the activities and the data captured by the two watches are aligned and synchronized.

Once the data is recorded, we load the data from the different sensors. In case of using the raw data from the IMU (accelerometer, gyroscope, magnetometer and attitude), the angular acceleration has to be transformed to linear one using quaternions (for instance with the AHRS algorithm\footnote{\url{https://x-io.co.uk/open-source-imu-and-ahrs-algorithms/}}), obtaining also the Euler angles. Since the Apple Watch also gives the linear acceleration, we do not need to convert the angular to linear acceleration. Thus, we only have to preprocess the acceleration to minimize the sensor drift, which often leads to inaccurate measures.



\subsection{Data labelling }\label{s:prep}
We have manually labeled the captured data for training the classification and gesture spotting approach. Thus, we isolate all movements and label them with the corresponding movement ($\mathcal{M}_i$ and $\mathcal{R}_j$). In this way, with the groundtruth timestamps from the user recordings, we can get the exact positions of the target movements and the time where the user was resting calm.

\subsection{Segmentation}

The segmentation step aims to obtain the subsequences that are candidates of being a target movement. These subsequences will be later classifed, whereas the rest of the sequence will be discarded. In order to detect a target movement, we explored two options: considering the entire action or only a part of it (gesture). 
Since L1 is a constraint scenario, it is easier to segmented because users make a pause between movements. Contrary, L2 is an unconstrained continuous stream signal, so it is more difficult to automatically segment given that L2 was designed to simulate real life conditions. Consequently, the segmentation is held differently in each scenario, as described next.

\subsubsection{Scenario L1}
In the constrained continuous scenario L1, we use the following segmentation options:
\begin{itemize}
    \item \emph{Action segmentation.} In this case, the sequences in L1 are segmented thanks to the very short rest time between the sequence of target movements. So, whenever an inappreciable movement is recorded by the sensors, the sequence is segmented.
    
    \item \emph{Gesture spotting.} In order to speed-up the detection and classification time, we propose gesture spotting. Since the peak of the signal is widely employed as a classification feature in activity recognition~\cite{max1}~\cite{max2}, we also explore this possibility. Thus, instead of classifying the entire action, we only segment the relevant parts of the action. In our case, the relevant part is the positive peak and a small part of the motionless linear acceleration signal before and after that peak, as shown in Figure~\ref{fig: gest}.
\end{itemize} 

\subsubsection{Scenario L2}
\begin{figure}[h]
\centering
\includegraphics[width=0.7\columnwidth]{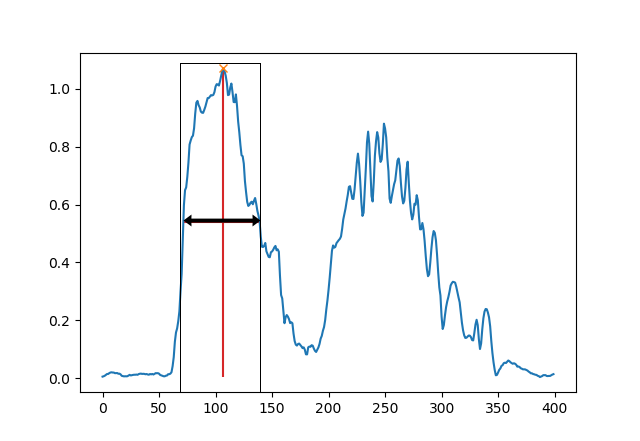} 
\caption{Gesture Spotting illustration. The peak is shown in a vertical red line. The relevant part is the region covered by the rectangular bounding box shown in black color.}
\label{fig: gest}
\end{figure}

In the unconstrained continuous scenario L2, we opt for these two segmentation options:
\begin{itemize}
    \item \emph{Action segmentation with Non-overlapping Sliding windows.} Sliding windows have been traditionally used to exhaustively analyse sequential data, although they imply a high computational cost. Sliding windows are commonly used~\cite{sliding_window} in two forms: overlapping or non-overlapping windows. After various experiments, we experienced that non-overlapping windows are preferable. The optimal "size" of the sliding window had been experimentally set using the training data.
    \item \emph{Gesture spotting.} In this scenario we also try to speed-up the detection using gesture spotting. Thus, as in L1, we segment the part around the positive peak of the sliding window, as shown in Figure~~\ref{fig: gest}.
\end{itemize}

\subsection{Classification}

Given the particularities and the few available labelled data, we have explored two different classification methods. The first one is a classical machine learning approach (SVMs), whereas the second one is a deep learning model (CNN).

\subsubsection{Support Vector Machines (SVM).} This first choice is motivated by the fact that SVM perform well in small datasets~\cite{med1}. In addition, it has also been reported that SVMs are frequently used in classification medical task:  decision-making, estimation of drug synergy, therapy synergy~\cite{med2}.
As explained in section \ref{s:SoA}, Support Vector Machines have been typically used in HAR because of their efficiency in data classification and classes separation. The Apple Watch provides the following information: acceleration, rotation, yaw, pitch, roll. For classification, we do take into account all the provided sensors' information. 
In our case, we have evaluated different sets of feature vectors, and we have experimentally found that the most suitable minimalist feature set is the mean, the minimum, the maximum and the standard variation of a window.

\subsubsection{Convolutional Neural Networks (CNN).}
The typical signal classification pipelines usually start with a pre-processing step, and subsequently, a feature extraction stage. Obviously, a good choice of the feature descriptors is important to avoid omitting relevant signal features that could affect the classification. So, to palliate the above mentioned issue, and contrary to the SVMs approach described above that uses a defined feature vector set, we alternatively opt to use the preprocessed raw signal as the input of the Convolutional Neural Networks model. In this classification model, we use all the time points of the window or gesture as input. 

\begin{figure*}[h]
\centering
\includegraphics[width=0.9\textwidth]{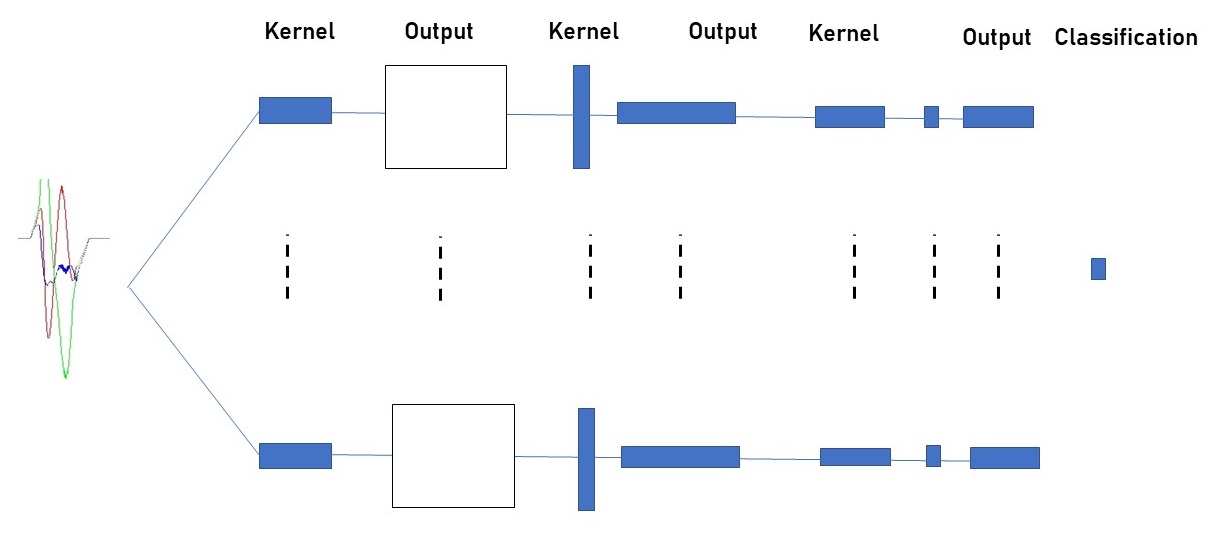}
\caption{CNN-based model classification architecture.}
\label{fig_model}
\end{figure*}

We propose a CNN model inspired from EEGNet~\cite{eeg}, a compact CNN architecture intended to classify and interpret electroencephalography-based brain computer interfaces. The original architecture has been modified (concretely, the convolution dimensions) because the signal nature is different, both in terms of frequency and length. 
The input of in EEG is defined by (C,T), being C the number of channels and T the number of time points. Both C and T change in our case, since the recording frequency is 100Hz instead 128Hz and the number of channels is 12.
Accordingly, the first filter convolution size is set to be half of the sampling rate (50 in our case).
In the first part of the architecture, two convolutions are carried out, in sequence. Next, we have a wise separable convolution, so that we reduce the number of parameters and computations while scaling up representational efficiency. Finally, the resulted features are passed to a softmax for the final classification. The proposed architecture is shown in Figure~\ref{fig_model}.

\section{Results} \label{s:Results}

In this section we describe the dataset and discuss the experimental results.

\subsection{Dataset}

\begin{figure}[h]
\centering
\begin{tabular}{cc}
\includegraphics[width=0.45\columnwidth]{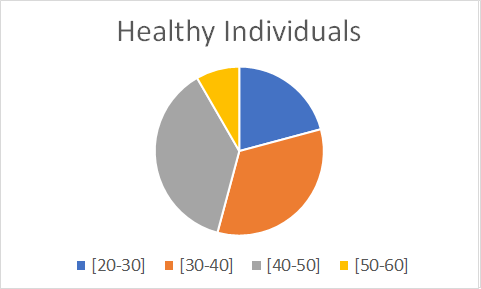} &
\includegraphics[width=0.45\columnwidth]{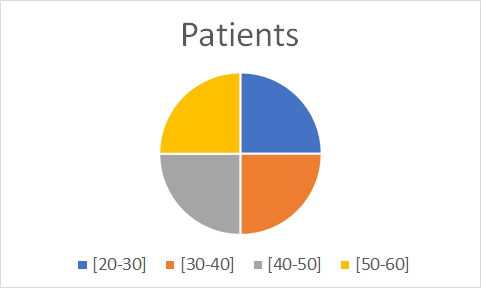}\\
(a) & (b)\\
\end{tabular}
\caption{Population statistics of healthy individuals (a) and patients (b).}
\label{fig_age}
\end{figure}

The dataset is collected using the Apple Watch integrate sensors. We have recorded 25 healthy subjects and 4 patients in the L1 and L2 scenarios described before. The healthy population's age distribution is shown in Figure~\ref{fig_age}a). The gender percentages are 48\% women, 52\% men. Concerning the patients, there are 3 men and 1 woman. The patients' age distribution is shown in Figure~\ref{fig_age}b).
This amount and distribution of users (in terms of age and gender) aims to provide enough variation in the performing style of each movement, and thus, ease the training of the classification algorithms.

In scenario L1, each user is recorded multiple times while performing each one of the four key target movements [$\mathcal{M}_1$....$\mathcal{M}_4$]. Afterwards, the user performs three L2 sessions. An L2 session consists of a random sequence of target $\mathcal{M}_i$ and non-target $\mathcal{R}_j$ actions. Obviously, all sessions are different since the sequence movements is randomly selected. This means that no user repeats the same sequence of movements. 

The dataset will be available the project's website \footnote{Dataset available 
at \url{http://dag.cvc.uab.es/patientmonitoring/}}.

\subsection{Results}
The baseline results are presented in this subsection. For the CNN classifier, we randomly split our data into 60\% for training, 20\% for validation and 20\% for testing. In the case of the SVM classifier, we use the same 20\% for testing, whereas the remaining 80\% is used for training (no validation set).

\begin{table}[t]
\caption{Classification accuracies of the SVM and CNN-based classifiers in the test set. The higher the value, the better.}
\centering
\begin{tabular}{|l|l|l|l|l|l|l|} 
\hline
\multirow{3}{*}{Scenario} & \multicolumn{4}{l|}{Healthy subjects}                               & \multicolumn{2}{l|}{Patients}  \\ 
\cline{2-7} & \multicolumn{2}{l|}{Action Segm.} & \multicolumn{2}{l|}{Gesture Spotting} & Action Segm. & Gesture Spotting            \\ 
\cline{2-7}
                          & SVM  & CNN                  & SVM  & CNN                   & SVM     & SVM                 \\ 
\hline
L1                        & 84\% & 65\%                 & 55\% & 60\%                  & 56\%    & 41\%                \\ 
\hline
L2                        & 61\% & 59\%                 & 51\% & 53\%                  & 41\%    & 35\%                \\
\hline
\end{tabular}
\label{tab:results}
\end{table}

The performance results of the two approaches are shown in Table \ref{tab:results}. Concerning the evaluation of healthy individuals' data, we observe that, in general, the SVM classifier obtains better results. The SVM classifier reaches an accuracy of 84\% in L1. However, the random sequences and the size of the NULL class in the L2 scenario makes it hard to achieve similar classification results, so the SVM accuracy decreases to 61\%. The CNN classifier obtains lower results than the SVM classifier. We believe that the small size of the dataset plays a major role in the lessening of the performance, because deep learning methods usually need more training data than classical machine learning approaches. As in the SVM classifier, the CNN's accuracy is slightly lower in the L2 scenario. 

Regarding gesture spotting, we observe that classifying the entire action using sliding windows obtains better accuracies. However, the classification via gesture spotting highly reduces both the number of signals and also the length of the signal to evaluate. This suggests that it is more suitable for real-time applications running in smartwatches. Anyway, it must be noted that when classifying via gesture spotting, the results obtained by the SVM and CNN classifiers are quite similar, with a difference of 5 points in the L1 scenario (55\% versus 60\%) and only 2 points in the L2 scenario (51\% versus 53\%).

Concerning the evaluation of the patients' data, and given the few amount of training data and the results obtained with healthy subjects, here we only present the results related to the SVMs classifiers. We can notice a decrease in classification accuracy by more than 20\%, in scenario L1, compared to healthy individuals. The gesture spotting reaches only 35\% in the L2 scenario. The main reason behind this performance decrease is the fact that the patients experience hemiparesis (weakness of one side of the body), so the target movements are poorly performed. In consequence, it is extremely difficult to detect these target movements in the affected arm, especially in the first weeks of rehabilitation.

These results suggest that, given the difficulties in spotting the target movements in the impaired arm in patients, whenever the movements are symmetric (performed by the two arms at the same time), the gesture spotting might be based on the healthy arm solely.


\section{Conclusion} \label{s:Conclusions}


In this work we have proposed an upper-limb assessment framework for assessing the neuromotor status of stroke patients. This application protocol is particularly designed for unconstrained scenarios and based on non-repetitive movements inspired on the Fugl-Meyer scale, with the aim to simulate more realistic evaluation scenarios. We have constructed an experimental database consisting of gesture recordings of healthy subjects and stroke patients, with the corresponding ground truth. In addition of this protocol and dataset, we have also proposed an Activity Recognition baseline (SVMs and CNNs). We do expect that this protocol, dataset and baseline results will foster the research in the rehabilitation assessment field. 

Future work will focus on exploring data augmentation techniques for increasing the few available training data as well as transfer learning techniques for benefiting from similar HAR datasets. In the near future, we plan to integrate the spotting method in the full motor assessment pipeline, so that the automatic segmentation of the target movements will be the input of the kinematic analysis algorithm.


\section*{Acknowledgment}

This work has been partially supported by the H2020 ATTRACT EU project (Grant Agreement 777222, TPPA 773, RPM3D), the Spanish project RTI2018-095645-B-C21, the FI fellowship AGAUR 2020 FI-SDUR 00497 (with the support of the Secretaria d’Universitats i Recerca of the Generalitat de Catalunya and the Fons Social Europeu), the Ramon y Cajal Fellowship RYC-2014-16831 and the CERCA Program/\ Generalitat de Catalunya. 
C. Carmona-Duarte was supported by a Viera y Clavijo contract from the Universidad de Las Palmas de Gran Canaria. The authors would like to thank Oriol Ramos and \'Angel S\'anchez for fruitful discussions.



%


\bibliographystyle{unsrt}
\bibliography{bib.bib}

\end{document}